**Title:** Mechanical-Diode based Ultrasonic Atomic Force Microscopies

**Abstract**

Recent advances in mechanical-diode based ultrasonic force microscopy techniques are reviewed. The potential of Ultrasonic Force Microscopy (UFM) for the study of material elastic properties is explained in detail. Advantages of the application of UFM in nanofabrication are discussed. Mechanical-Diode Ultrasonic Friction Force Microscopy (MD-UFFM) is introduced, and compared with Lateral Acoustic Force Microscopy (LAFM) and Torsional Resonance (TR) - Atomic Force Microscopy (AFM). MD-UFFM provides a new method for the study of shear elasticity, viscoelasticity and tribological properties on the nanoscale. The excitation of beats at nanocontacs and the implementation of Heterodyne Force Microscopy (HFM) are described. HFM introduces a very interesting procedure to take advantage of the time resolution inherent in high-frequency actuation.

**Keywords**

**Table of contents**

**Author: M. Teresa Cuberes**

**Mailing Address:**

Laboratorio de Nanotécnicas, UCLM

Plaza Manuel de Meca 1, 13400 Almadén, Spain

**Tfn.** 34 902204100 ext. 6045 (office) 6038 (lab)

**Fax.** 34 926264401

**Email:** teresa.cuberes@uclm.es

# 1. Introduction: Acoustic Microscopy in the Near Field.

## 1.1 Acoustic Microscopy: possibilities and limitations

Acoustic microscopy uses acoustic waves for observation in a similar way as optical microscopy uses light waves. In acoustic microscopy, a sample is imaged by ultrasound, and the contrast is related to the spatial distribution of the mechanical properties. The procedure can be implemented in transmission as well as in reflection. The first Scanning Acoustic Microscope (SAM) was introduced in 1974 [1], and was mechanically driven and operated in the transmission mode. Nowadays most commercial acoustic microscopes work in the reflection mode; by using pulsed acoustic systems the reflections of the acoustic beam from the specimen may be separated from spurious reflections. Applications of Acoustic Microscopy [1-10] include mapping of inhomogeneities in density and stiffness in materials, measurement of coating thicknesses, detection of delaminations in electronic integrated circuit chips, detection of microcracks and microporosity in ceramics, identification of the grain structure and anisotropy in metals and composites, and evaluation of elastic properties of living cells, etc.

A schema of a SAM operating in reflexion mode is shown in Fig. 1. The acoustic microscope works on the principle of propagation and reflexion of acoustic waves at interfaces where there is a change of acoustic impedance ($Z = density \times velocity$).

Acoustic waves initiated at the piezoelectric transducer refract at the lens/coupling medium interface and focus to a diffraction limited spot. The sound wave is propagated to the sample through a couplant, usually water. When a sudden change in acoustic impedance is encountered, like at a material boundary, a portion of the sound energy is reflected, and the remainder propagates through the boundary. The transducer detects and converts the reflected acoustic waves into an electrical signal, which is digitized and stored at appropriate points during scanning of the lens. The amplitude and phase of the reflected acoustic signal determine the contrast in the acoustic images.

SAM also permits the implementation of time of flight (TOF) measurements. In time-resolved acoustic microscopy a short sound pulse is sent towards a sample. The time-of-flight method monitors the time required for the pulse sent into the sample to return back to the acoustic lens. TOF images provide a means to determine relative depth variations in the location of inhomogeneous or defective sites within a sample.

The spatial resolution and depth of penetration in SAM are inter-related, and dependent on the operating frequency. When using acoustic waves of frequencies around 1-2 GHz, SAM images with conventional optical resolution of the order of microns can be obtained. Nevertheless, when using low frequency ultrasound, in the 2-10 MHz range, the spatial resolution is typically limited to the millimetre range. On the contrary, the depth of penetration decreases as the frequency increases. In technical materials, the attenuation of pressure waves is given by the microstructure, and increases at least with the square of frequency. In fine-grained or fine-structured materials, the absorption, which increases linearly with frequency, limits the penetration. For GHz frequencies, the penetration depth may be of the order of microns. When the frequencies are in the MHz range the penetration may be in the millimetre range. The pressure waves, regardless of frequency, are more heavily attenuated in air than in liquids, so water is usually used as a convenient couplant between the transmitter/receiver of acoustic waves and the specimen.

Many advanced techniques based on SAM have emerged. In phase sensitive acoustic microscopy (PSAM) [7-9], phase and amplitude SAM images are simultaneously obtained. The phase detection mode can be implemented in transmission, and it permits the observation of propagating waves emitted from an acoustic lens in a holographic manner, as shown in Fig. 2. Hybrid SAM-based technologies, such as photoacoustic microscopy (PAM) have proved to be of extreme value (see for instance [10]).

In spite of the advantages of the aforementioned acoustic techniques, the resolution achievable when using acoustic waves for observation at its best is still poor for applications in nanotechnology. Acoustic imaging with nanometer scale resolution can be realized using Ultrasonic Atomic Force Microscopy techniques, as described in this chapter.

**1.2 Ultrasonic Atomic Force Microscopies**

The main motivation for the initial development of Ultrasonic Atomic Force Microscopies was to implement a near-field approach that provided information such as that obtained with the Acoustic Microscope, but with a lateral resolution on the nanometer scale. The area of mechanical contact between the tip of an atomic force microscope (AFM) cantilever and a sample surface is typically of the order of nanometers in diameter. An AFM cantilever with the tip in contact with a sample surface follows small-amplitude out-of-plane surface vibration linearly, provided its frequency is below the cantilever resonance frequency. One might expect that the cantilever would react the same to the pressure exerted at the tip-sample contact by an acoustic wave of millimetre wavelength, realizing ultrasound detection in the near-field. However, due to the inertia of the cantilever, the linear behaviour is not evident in the limit of high-frequency signals. As a matter of fact, if the cantilever is regarded as a simple point mass, the amplitude of vibration at the driving frequency vanishes in the limit of very high frequencies.

Basically, we may distinguish two different procedures for the detection of high-frequency surface mechanical vibration with the tip of an AFM cantilever. The first is based on the fact that actually the cantilever is not a point mass, but a tiny elastic beam that can support high-frequency resonant modes [11, 12]. When a cantilever tip is in contact with the sample surface and high-frequency surface vibration is excited at the tip-sample contact, the so-called *contact resonances* of the cantilever are excited at certain characteristic frequencies. Those depend on both the cantilever and the sample elastic properties [11-14]. Techniques such as Acoustic Atomic Force Microscopy (AFAM) [11] and Ultrasonic Friction Force Microscopy (UFFM) [15-17] are based on the study of cantilever contact resonances. Scanning Microdeformation Microscopy (SMM) [18, 19] and Scanning Local Acceleration Microscopy (SLAM) [20] also monitor the vibration of an AFM cantilever with the tip in contact with the sample surface at the ultrasonic excitation frequency. A second approach is based on the so-called *mechanical-diode* effect [21, 22], which will be explained in more detail in section 2.1. In this case, the operating ultrasonic frequency is extremely high, or so that the contact resonances of the cantilever are not excited. Then, the cantilever does not follow the high-frequency surface vibration due to its inertia. Nevertheless, if the surface ultrasonic vibration amplitude is sufficiently high that the tip-sample distance is varied over the non-linear regime of the tip-sample interaction force, the cantilever experiences an additional force, or *ultrasonic force*. This can be understood as the averaged force acting upon the tip in each ultrasonic cycle. As a result of the ultrasonic force, the tip experiences a displacement –ultrasonic displacement, or mechanical-diode response - that can be monitored, and which carries information about the elastic properties of the sample, and the adhesive properties of the tip-sample contact [22].

Techniques such as Scanning Acoustic Force Microscopy (SAFM) [23, 24], Ultrasonic Force Microscopy (UFM) [25], Mechanical-Diode Ultrasonic Friction Force Microscopy (MD-UFFM) [26], and Heterodyne Force Microscopy (HFM) [27] utilize mechanical-diode type responses.

Up-to-now, most ultrasonic-AFM studies have been performed with the tip in contact with the sample surface, although in principle non-contact ultrasonic-AFM techniques can be implemented using either the high-order cantilever resonance frequencies, or the mechanical diode effect, as long as the distance between the tip of an inertial cantilever and a sample surface is swept over a non-linear interaction regime. Recently, the possibility to detect high-frequency vibration using dynamic force microscopy has been demonstrated [28]; the detection of acoustic vibration in this case is apparently also facilitated because of the activation of the mechanical diode effect (see section 2.1 for further discussions).

This chapter is mostly devoted to review the fundamentals and recent advances in mechanical-diode based ultrasonic force microscopy techniques implemented in contact mode. In section 2, Ultrasonic Force Microscopy (UFM) will be explained in detail. The use of UFM in nanofabrication provides unique advantages [29]. In section 3, Mechanical-Diode Ultrasonic Friction Force Microscopy (MD-UFFM) [26] will be introduced. MD-UFFM is based on the detection of shear ultrasonic vibration at a sample surface via the lateral mechanical-diode effect. This is a new method for the study of shear elasticity, viscoelasticity and tribological properties on the nanoscale. Section 4 discusses the technique of Heterodyne Force Microscopy (HFM) [28]. HFM provides a novel and very interesting procedure to take advantage of the time resolution inherent in high-frequency actuation. In HFM, mechanical vibration in the form of beats is induced at the tip-sample contact by simultaneously launching ultrasonic waves towards the tip-sample contact region from the cantilever base and from the back of the sample, at slightly different frequencies. If the launched cantilever and sample vibration amplitudes are such that the tip-sample distance is varied over the non-linear tip-sample force regime, the cantilever vibrates additionally at the beat frequency due to the mechanical-diode effect (beat effect). HFM monitors the cantilever vibration at the beat frequency in amplitude and phase. As has been demonstrated, Phase-HFM provides information about dynamic relaxation processes related to adhesion hysteresis at nanoscale contacts with high time sensitivity [28]. Recently, Scanning Near-Field Ultrasound Holography (SNFUH) [30] has been introduced. The principle of operation is very similar to that of HFM. The experimental data reported by SNFUH demonstrate its capability to provide elastic information of buried features with great sensitivity. Also, the technique of Resonant Difference-Frequency Atomic Force Ultrasonic Microscopy (RDF-AFUM) [31] has been proposed, based on the beat effect. Discussions of the beat effect and HFM will be included in section 4.

## 2. Ultrasonic Force Microscopy (UFM): The Mechanical Diode Effect.

### 2.1 The mechanical diode effect

The first observation that the tip of an AFM cantilever can be used to detect out-of-plane high frequency vibration of a sample surface was reported in [21]. In these experiments, Surface Acoustic Waves (SAWs) were excited at (slightly) different frequencies by means of interdigital transducers (IDTs) and the frequency of surface vibration was detected using a cantilever tip in contact with the sample surface. The technique of Scanning Acoustic Force Microscopy (SAFM) has been demonstrated for the characterization of SAWs field amplitudes [24] and phase velocities [32]. Acoustic fields in bulk acoustic-wave thin-film resonators have also been imaged with this method [33].

The physical mechanism that allows a cantilever to detect out-of-plane surface ultrasonic vibration excited at the tip-sample contact is based on the nonlinearity of the tip-sample interaction force [22]. Even though it is expected that inertia prevents a cantilever tip in contact with a sample surface to move fast enough to keep up with surface atomic vibrations at ultrasonic frequencies, the displacement of the surface leads to modification of the tip-sample force $F_{t\text{-}s}$ provided the ultrasonic vibration amplitude is sufficiently high and the tip-sample distance $d$ is varied over the nonlinear tip-sample force regime. In Fig. 3, it is assumed that the tip is in contact with a sample surface, in the repulsive force regime. When out-of-plane surface ultrasonic vibration is switched on the tip-sample distance $d$ is varied at ultrasonic frequencies between some minimum and maximum values, corresponding to the amplitude of ultrasound excitation. If the ultrasonic amplitude of is small, the tip-sample distance sweeps a linear part of the tip-sample interaction force curve. In this case, the net averaged force that acts upon the cantilever during an ultrasonic time period is equal to the initial set-point force, and hence the deflection of the cantilever remains the same as in the absence of ultrasound.

However, if the amplitude of ultrasonic vibration is increased, the tip-sample distance sweeps over the nonlinear part of the force curve, and the averaged force includes an additional force $F_{ult}$ given by

$$F_{ult}(d) = \frac{1}{T_{ult}} \int_0^{T_{ult}} F_{t-s}\left(d - a \cdot \cos\left(\frac{2\pi}{T_{ult}}t\right)\right) dt \qquad (eq.\,1)$$

where $d$ the tip-sample distance, $a$ is the amplitude of ultrasonic vibration, and $T_{ult}$ the ultrasonic period.

Due to this additional force, named hereafter the *ultrasonic force*, the cantilever experiences an additional deflection which can be easily detected by means of the optical lever technique, and is the physical parameter which is monitored in Ultrasonic Force Microscopy (UFM) [25]. The *UFM deflection* is a quasi-static cantilever deflection that occurs as long as out-of-plane ultrasonic vibration of sufficiently high amplitude is present at the tip-sample contact. The quasi-static equilibrium deflection is given by:

$$k_c z_{eq} = F_{ult}(d_{eq}, a) \qquad (eq.\,2)$$

where $k_c$ is the cantilever stiffness, and $z_{eq}$ and $d_{eq}$ are the new cantilever deflection and tip indentation depth respectively. As the surface ultrasonic vibration amplitude is further increased, $F_{ult}$ increases due to the nonlinearity of the tip-sample force curve, and hence the cantilever deflection increases too until a new equilibrium position is reached. In this sense, the cantilever behaves as a "mechanical diode" [20], and deflects when the tip-sample contact vibrates at ultrasonic frequencies of sufficiently high amplitude.

To perform UFM, the ultrasonic excitation signal is typically modulated in amplitude with a triangular or trapezoidal shape (see Fig. 4). In UFM, the ultrasonic amplitude modulation frequency is chosen to be much lower than the first cantilever resonance, but higher than the AFM feedback response frequency to avoid that the feedback compensates for the ultrasonic deflection of the cantilever. Hence, contact-mode AFM can be performed to obtain a surface topographic image, in spite of the presence of ultrasound. To record an UFM image, the ultrasonic deflection of the cantilever is tracked at the amplitude modulation frequency using a lock-in amplifier.

To detect surface ultrasonic vibration with dynamic force microscopy in ref. [28], a resonator is used as a sample, and its acoustic vibration (at about 1.5 GHz) is modulated in amplitude with a sinusoidal shape. The ultrasonic amplitude modulation frequency is chosen to be coincident with the second eigenmode of the cantilever. Typical Dynamic Force Microscopy is performed using the first eigenmode of the AFM cantilever (at about 72 KHz) in order to obtain a surface topographic image, which can be properly done in spite of the presence of acoustic vibration. To obtain acoustic information, the cantilever vibration in the second eigenmode (at about 478 KHz) is monitored with a lock-in amplifier. The surface acoustic vibration occurs in the GHz range, and the cantilever tip oscillates in the $10^2$ KHz range. Hence, it can be considered that at each point of the cantilever tip vibration cycle at the ultrasonic amplitude modulation frequency in the $10^2$ KHz range, the tip-sample distance varies many times, due to the sample vibration in the GHz range. The cantilever cannot vibrate at the resonator GHz frequencies due to its inertia. This results in a periodic ultrasonic force acting upon the cantilever, with a period corresponding to the ultrasonic amplitude modulation frequency, i.e. the second cantilever eigenmode.

**2.2 Experimental implementation of UFM**

The experimental set-up for UFM can be implemented by appropriately modifying a commercial AFM [25, 34]. A schema of an UFM apparatus is shown in Fig. 4.

An ultrasonic piezoelement is located on the sample stage and the sample is directly bonded to the piezo using a thin layer of crystalline salol, or just honey, to ensure good acoustic transmission. In this way, longitudinal acoustic waves may be launched from the back of the sample to the sample surface. A function generator is needed to excite the piezo and generate the acoustic signal (Fig 4). The ultrasonic deflection of the cantilever is monitored using the standard four-segment photodiode. As mentioned in section 2.1, the ultrasonic signal is modulated in amplitude with a triangular or trapezoidal shape, with a modulation frequency above the AFM feedback response frequency. The UFM response (ultrasonic deflection) can be monitored with a lock-in amplifier using the synchronous signal provided by the function generator at the ultrasonic modulation frequency. In this way, contact-mode AFM topographic images and UFM images can be simultaneously recorded over the same surface region.

In addition to the described configuration, it is also possible to perform UFM by exciting the ultrasonic vibration at the tip-sample contact using a piezotransducer located at the cantilever base [35, 36]. This latter procedure has been named Waveguide-UFM. In this case, the ultrasonic vibration is propagated through the cantilever to the sample surface. Here, the cantilever tip which should necessarily vibrate at the ultrasonic excitation frequency. However, if the ultrasonic frequency is sufficiently high, the amplitude of high-frequency cantilever vibration can be very small, and it has been experimentally demonstrated that a mechanical-diode cantilever response (i.e. an ultrasonic deflection of the cantilever) is activated well under these conditions [35, 36].

UFM responses are also detected in liquid environments [37]. To perform UFM in liquid, the ultrasonic piezoelectric transducer is simply attached with honey to the back of the sample-holder stage of the AFM liquid cell [37].

### 2.3 Information from UFM data

*UFM curves*

In order to study the UFM response, UFM data are typically collected in the form of *ultrasonic curves*, obtained at each surface point by monitoring the cantilever ultrasonic deflection or mechanical-diode response as a function of the ultrasonic excitation amplitude. In the following, a description of the current understanding of those curves is provided.

As discussed in section 2.1, the UFM signal stems from the time-averaged force exerted upon a cantilever tip in contact with a sample surface when ultrasonic vibration of sufficiently high amplitude is excited at the tip-sample contact, in such a way that the tip-sample distance is varied over the nonlinear tip-sample force regime at each ultrasonic period. The forces acting at a cantilever tip in contact with a sample surface are often described in the context of continuum mechanics. In particular, a tip-sample force – indentation curve with shape as depicted in Fig. 5 (a) can be derived from the Johnson-Kendall-Roberts (JKR) model that describes a sphere pressed against a flat surface. The pull-off distance is defined as the tip-sample distance at which the tip-sample contact breaks when the tip is withdrawn from the sample surface. If the tip-sample indentation is varied over the linear tip-sample force regime, as it is the case for the amplitude $a_o$ in Fig. 5 (a), the average force is $F_i$. If the vibration amplitude is $a_1$, the pull-off point is reached and the tip-sample contact is broken for a part of the ultrasonic cycle; the ultrasonic curve -cantilever deflection versus ultrasonic amplitude- shows a discontinuity at this amplitude value. Fig. 5 (b) schematically shows the ultrasonic deflection (UFM signal) that will be received as the surface ultrasonic vibration amplitude is linearly increased. The mechanical diode response or ultrasonic cantilever deflection experiences a discontinuity attributed to an *ultrasonic force jump* when the vibration amplitude reaches the so-called *threshold amplitude* $a_1$.

Fig. 6 displays experimental UFM curves recorded on highly oriented pyrolitic graphite (HOPG) when the surface ultrasonic vibration amplitude is varied with a trapezoidal shape as indicated, lowest curve in the figure, for various initial loads. Notice that, as expected, when the initial tip-sample force is increased the ultrasonic thresholds occur at higher vibration amplitudes. The ultrasonic deflection of the cantilever is dependent on both the initial set-point force and the ultrasonic excitation amplitude. For a given set-point force, the threshold amplitude is needed to reach the pull-off point and induce the jump of the ultrasonic cantilever deflection. Consistently, the threshold amplitude increases as the set-point force is increased. From the analysis of the ultrasonic curves, information about the tip-sample interaction force can be obtained, and the elastic and adhesive properties of the tip-sample contact can be derived.

The procedure of *differential UFM* has been proposed to extract quantitative information about the sample stiffness with nanoscale resolution [38], based on the measurement of the threshold amplitudes $a_i$ of the ultrasonic curves for two different initial tip-sample normal forces $F_i$. If the normal forces do not differ much, the effective contact stiffness $S_{eff}$ can be obtained as follows

$$S_{eff}(F_{av}) = \frac{F_2 - F_1}{a_2 - a_1} \qquad (eq.3)$$

$$F_{av} = \frac{F_2 + F_1}{2} \qquad (eq.4)$$

This method has the advantage that for the derivation of the contact stiffness it is not necessary to consider the details of a specific contact-mechanics model for the tip-sample interaction.

Simulations of the UFM curves have been done introducing the concept of *modified tip-sample force curves* [38] (see Fig. 7). When the tip-sample distance is varied because of the excitation of ultrasound, the tip-sample interaction forces are modified because of the mechanical-diode effect. In order to simulate the UFM curves, a series of modified tip-sample force curves are generated, each of them corresponding to a specific value of the ultrasonic amplitude. In Fig. 7, force curves obtained for ultrasonic vibration amplitudes $a_o$ and $a_1$ ($a_o<a_1$) have been plotted together with the original force curve in the absence of ultrasonic vibration, derived from the JKR model. The straight line in Fig. 7 represents the Hooke law, which relates the force acting on the tip to the cantilever normal deflection. Here, the cantilever tip is modelled as a point mass on a spring. The equilibrium positions of the tip in contact with the surface are obtained from the intersection of the line with the corresponding force curve, which varies depending on the ultrasonic excitation amplitude. It can be noticed that the pull-off forces and the indentation values are modified at the new modified force curves; for a given indentation value, the new force value is generally higher than the one obtained for zero ultrasonic amplitude. It may also be noted that for the amplitude $a_1$, which corresponds to the threshold amplitude, there are two solutions for the new equilibrium position of the tip, which accounts for the discontinuity in the cantilever displacement or force jump at this amplitude value.

The stiffness values chosen to generate the original JKR curve can be derived from the application of differential UFM to the simulated ultrasonic curves, giving confidence in the reliability of this method [38]. From the analysis of the dependence of the simulated UFM curves on the sample Young modulus and adhesion, it can be concluded that (i) the threshold amplitude increases when the normal force is increased, the Young modulus is low or the work of adhesion is high, and (ii) the force jump increases when the Young modulus is low and the work of adhesion is high [34]. Analysis of the ultrasonic curves with other contact models for the tip-sample interaction yield similar conclusions [39].

Fig. 6 shows that when the ultrasonic amplitude at the tip-sample contact is linearly decreased, the cantilever returns to its original equilibrium position, experiencing a sudden jump-in in the force. The ultrasonic amplitude at which the jump-off in the UFM response is observed when increasing the excitation amplitudes, i.e. the threshold amplitude, is different from that at which the jump-in occurs when the amplitude is decreased A method has been proposed [40] to determine both the sample elastic modulus and the work of adhesion from such force jumps in the ultrasonic curves. In [41, 42] the area between experimental ultrasonic curves obtained increasing and decreasing the ultrasonic amplitude - due to the different jump–off and jump-in threshold amplitudes- is defined as the *UFM hysteresis area* (UH), and related to the local adhesion hysteresis. Correlations between the adhesion hysteresis and the local friction were theoretically and experimentally investigated [43, 44]. Using the ability of UFM to provide information about local adhesion hysteresis, the protein-water binding capacity was investigated in protein films at different relative humidities, with the proteins in hydrated and dehydrated states [45].

The transfer of ultrasound to an AFM cantilever in contact with a sample surface has also been evaluated by numerically solving the equation of motion, taking into account the full nonlinear force curve and considering that the cantilever is a rectangular beam that supports flexural vibrations. By this procedure, the change in the mean cantilever position that results from the nonlinear tip-sample interactions is also demonstrated [46].

*UFM images*

The ability of UFM to map material properties simultaneously with the acquisition of contact-mode AFM topographic images over the same surface area has been extensively demonstrated. Given the set-point force and the maximum ultrasonic amplitude, if assumed that the tip-sample adhesion is invariant, the UFM signal corresponding to a locally stiff region is large in magnitude, and gives rise to a bright contrast in the UFM image. So far the UFM signal depends on both adhesion and elasticity, the contrast in the UFM images must be carefully analyzed. The UFM brings advantages for the study of both soft and hard materials. In the presence of surface out-of-plane ultrasonic vibration of sufficiently high amplitude, nanoscale friction reduces or vanishes [47, 48], which facilitates the inspection of soft samples without damage. The elastic properties of hard materials can also be investigated by UFM. Due to the inertia of the cantilever, in the presence of surface ultrasonic vibration a cantilever tip effectively indents hard samples [25, 20].

Important applications of UFM rely on its capability to provide subsurface information. The subsurface sensitivity of the UFM has been experimentally demonstrated [25, 20]. Subsurface dislocations in HOPG have been observed and manipulated (see Fig. 8) using the ultrasonic AFM [25, 49, 50].

The penetration depth in AFM with ultrasound excitation is determined by the contact-stress field, which increases when the set-point force and the ultrasonic amplitude are increased. The penetration depth and the minimum detectable overlayer thickness in Atomic Force Acoustic Microscopy (AFAM) are defined in [51], on the basis of the detectable minimum contact stiffness change. In [52] it is concluded from the change in contact stiffness of $SiO_2$/Cu, that buried void defects ($\approx$ 500 nm) at nm distance from the dielectric surface can be detected using the UFM. In [20], a GaAs grating buried under a polymeric layer is clearly imaged using Scanning Local Acceleration Microscopy (SLAM). Changes in contact stiffness of cavities in Si within about 200 nm from the Si(100) surfaces have been detected using UFM [53]. In [30], Au particles with a diameter of 15-20 nm buried under a 500 nm polymeric film have been observed using Scanning Near-Field Ultrasound Holography (SNFUH). UFM has been applied to characterise defects such as debonding, delaminations, and material inhomogeneities [54-56]. Subsurface information is also apparent in Resonant Difference-Frequency Atomic Force Ultrasonic Microscopy (RDF-AFUM) [31]. The UFM has been used to map stiffness variations within individual nanostructures such as quantum dots [57] or nanoparticles [58].

**2.4 Applications of UFM in nanofabrication**

Ultrasonic AFM techniques provide a means to monitor ultrasonic vibration at the nanoscale, and open up novel opportunities to improve nanofabrication technologies [49, 59]. As discussed above, in the presence of ultrasonic vibration, the tip of a soft cantilever can dynamically indent hard samples due to its inertia. In addition, ultrasound reduces or even eliminates nanoscale friction [47, 48]. Typical top-down approaches that rely on the AFM are based on the use of a cantilever tip that acts as a plow or as an engraving tool. The ability of the AFM tip to respond inertially to ultrasonic vibration excited perpendicular to the sample surface and to indent hard samples may facilitate nanoscale machining of semiconductors or engineering ceramics in a reduced time.

Fig. 8 demonstrates the *machining of nanotrenches and holes* on a silicon sample in the presence of ultrasonic vibration. Interestingly, no debris is found in the proximity of lithographed areas. Fig. 8 (a) refers to results performed using a cantilever with nominal stiffness in the 28-91 N m$^{-1}$ range and a diamond-coated tip. Fig. 9 (b) refers to results achieved using a cantilever with nominal stiffness 0.11 Nm$^{-1}$ and a Si$_3$N$_4$ tip; in the absence of ultrasound, it was not possible to scratch the Si surface using such a soft cantilever. In the machining of soft materials, as for instance plastic coatings, the ultrasonic-induced reduction of nanoscale friction may permit eventual finer features and improved surface quality in quasi-static approaches.

In bottom-up approaches, ultrasound may assist in *self-assembly or AFM manipulation of nanostructures* [49]. Effects such as sonolubrication and acoustic levitation have been studied at the microscale. These phenomena may facilitate a tip-induced motion of nano-objects. In the manipulation of nanoparticles (NPs) on ultrasonically excited surfaces with the tip of an AFM cantilever, both the tip-particle and particle-surface frictional properties change [59-61]. Moreover, the excitation of NP high-frequency internal vibration modes may also modify the NP dynamic response, and introduce novel mechanisms of particle motion. Using the UFM mode for manipulation allows us to monitor the mechanical diode response of the cantilever while individual nanoparticles are being laterally displaced over a surface by tip actuation, and receive information about the lateral forces exerted by the tip.

Eventually, it should be pointed out that the sensitivity of ultrasonic-AFM to subsurface features makes it feasible to monitor *subsurface modifications* [49]. We have recently demonstrated that actuation with an AFM tip, in the presence of ultrasonic vibration can produce stacking changes of extended grapheme layers, and induce permanent displacements of buried dislocations in Highly Oriented Pyrolytic Graphite (HOPG). This effect is illustrated in Fig. 10. In the presence of normal surface ultrasonic vibration, both AFM and lateral force microscopy (LFM) images reveal subsurface features [49, 59]. Subsurface modification was brought about in this case by scanning in contact mode, with high set-point forces, and high surface ultrasonic excitation amplitudes [49].

## 3. Mechanical Diode Ultrasonic Friction Force Microscopy (MD-UFFM)

### 3.1 The lateral mechanical diode effect

A lateral MD effect has also been experimentally observed. Similar to the UFM cantilever deflection that switches on in the presence of out-of-plane surface ultrasonic vibration of sufficiently high amplitude, an additional torsion of the cantilever is activated when the cantilever tip is in contact with a sample surface and scans laterally over the surface at low frequency. This is done in the presence of shear surface ultrasonic vibration of sufficiently high amplitude [24, 26].

The lateral MD-effect is exploited in Lateral Scanning Acoustic Force Microscopy (LFM-SAFM) [24] to obtain information about the amplitude and phase velocity of in-plane polarized SAWs. Recently, the technique of MD-UFFM has been proposed [26] to study the shear contact stiffness and frictional response of materials on the nanoscale. In MD-UFFM, shear ultrasonic vibration is excited at a tip-sample contact using a shear piezoelectric element attached to the back of the sample. Shear acoustic waves originated at the piezo propagate through the sample to reach the tip-surface contact area. An ultrasonic-induced additional torsion of the cantilever or MD-UFFM cantilever torsion is observed while the cantilever tip in contact with the surface is laterally scanning at low frequencies [26]. Experimental evidence of the lateral MD effect is provided in Fig. 10.

Fig. 10 (a) and (b) show typical MD-UFFM cantilever responses recorded on a Si sample in forward and backward scans respectively, in the presence of shear ultrasonic vibration at the tip sample-contact modulated in amplitude with a triangular shape. In both scanning directions, the ultrasound-induced torsion of the cantilever diminishes initially due to friction. As the shear ultrasonic excitation amplitude is increased, the MD-UFFM cantilever torsion increases in magnitude until a critical shear ultrasonic amplitude is reached, after which it remains invariant or decreases.

The lateral MD effect can be understood by considering the lateral ultrasonic force emerging from interaction of the tip with the lateral surface sample potential [26].

Fig. 11 illustrates a physical explanation for the MD effect, in agreement with experimental results [26]. In the absence of ultrasound, when scanning at low velocity on a flat surface, the cantilever is subjected to an initial torsion due to friction. At the typical low AFM scanning velocities, nanoscale friction proceeds by the so-called stick-slip mechanism [62]. At a sticking point, the tip is located at a minimum of the sum of the periodic surface potential and the elastic potential of the cantilever; the lateral displacement of the cantilever support relative to the sample introduces an asymmetry in the total potential that facilitates the jumping of the tip to the next energy minimum site. Most of the time, the tip sticks to a surface point, and then slips to a next sticking point with some energy dissipation. In Fig. 11, *E* corresponds to the total potential acting upon the tip when scanning forward at low velocity. Due to this potential, the tip is subjected to the force given by the derivative curve. When the tip lies in the minimum energy site crossed by the dashed line, the corresponding force is zero. Due to the different time-scales, we may consider that the tip-sample potential brought about by scanning at low velocity is frozen during a shear ultrasonic vibration period. The shear ultrasonic wave transmitted through the sample introduces in-plane oscillations at the sample surface, in the direction perpendicular to the long cantilever axis. Atomic species within the tip-sample contact area is subjected to shear ultrasonic vibration, but the inertia of the cantilever hinders its out-of-resonance rotation. The lateral displacement of the surface atoms relative to the tip leads to a time-dependent variation of the total potential acting upon the tip at ultrasonic time scales. We define the lateral ultrasonic force as the average force that acts upon the cantilever during each ultrasonic cycle in the presence of shear ultrasonic vibration,

$$F_{ult}^{l}(x_l, A) = \frac{1}{T_{ult}} \int_0^{T_{ult}} F\left(x_l - A * \cos\left(\frac{2\pi}{T_{ult}} t\right)\right) dt \qquad (eq. 5)$$

where $x_l$ is the lateral equilibrium location of the tip in the presence of lateral ultrasonic vibration of amplitude *A*, which defines the new equilibrium torsion of the cantilever, *A* is the amplitude of shear ultrasonic vibration and $T_{ult}$ refers to the ultrasonic time period. Once a critical lateral vibration amplitude is reached, sliding sets in, and the MD-UFFM signal does not increase further. A study of the MD-UFFM cantilever torsion may provide information about the sample shear stiffness and frictional response.

In Fig. 10, a vertical lift-off of the cantilever or *MD-UFFM cantilever deflection* is observed as a result of the shear surface ultrasonic vibration. Samples such as silicon are known to be covered by a liquid layer under ambient conditions. In such samples, the observed lift-off may originate from an elastohydrodynamic response of an ultrathin viscous layer sheared at the tip-sample contact at ultrasonic velocities [15]. The study of the MD-UFFM cantilever deflection may provide information about the elastohydrodynamic properties of thin confined lubricant layers.

**3.2 Experimental implementation of MD-UFFM**

The experimental set-up for Mechanical Diode - Ultrasonic Friction Force Microscopy (MD-UFFM) measurements can be implemented by appropriately modifying a commercial AFM [26]. The set-up required for MD-UFFM is similar to that required for the UFM substituting the longitudinal ultrasonic piezoelectric transducer with a shear-wave type. The shear-wave piezotransducer is mounted below the sample with its polarization perpendicular to the longitudinal axis of the cantilever. The sample should be attached to the sample with an appropriate couplant, as for instance crystalline salol. The changes in the cantilever torsion due to the lateral MD effect can be monitored in both forward and backward scans using the laser deflection method with a standard four-segment photodiode, simultaneously with the acquisition of contact-mode topographic images. MD-UFFM images can be collected by modulating the amplitude of the shear ultrasonic excitation and using a lock-in amplifier to detect the MD-UFFM signal. We distinguish torsion and deflection MD-UFFM modes, depending on whether the shear-ultrasonic-vibration-induced cantilever torsion or deflection response is studied.

In shear-wave piezoelectric transducer, parasitic out-of-plane vibration may arise due to the existence of bounderies, etc. In the presence of out-of-plane ultrasonic vibration of sufficiently high amplitude, the normal mechanical diode effect described in section 2.1 would lead to the excitation of an additional out-of-place cantilever deflection related to the sample elastic properties. In the absence of out-of-plane ultrasonic vibration, but with shear ultrasonic vibration excited on the sample surface, a lift-off or deflection of the cantilever is expected as a result of elastohydrodynamic lubrication effects of ultrathin viscous layers compressed at the tip-sample contact [15, 26]. In order to distinguish between those two effects, the UFM response of the sample under study and the used shear-wave piezotransducer should be very well characterized before establishing definitive conclusions from MD-UFFM measurements.

**3.3 Comparison of MD-UFFM with UFFM and TRmode AFM**

In UFFM, also named Lateral-Acoustic Friction Force Microscopy (L-AFAM) or Resonant Friction Force Microscopy (R-FFM)) [15-17, 64, 65] surface in-plane vibration polarized perpendicular to the long axis of the cantilever is excited with a shear-wave piezotransducer bonded to the back of the sample, as in MD-UFFM. UFFM monitors the torsional vibration of the cantilever at the sample shear ultrasonic excitation frequency, being the cantilever tip in contact with the sample surface. At shear ultrasonic frequencies, the torsional cantilever vibration is only significant near the cantilever torsional contact resonances. Fig. 12 shows UFFM measurements at a torsional contact resonance, cantilever torsional vibration amplitude versus surface shear ultrasonic excitation frequency, for different shear ultrasonic excitation amplitudes. At low shear excitation voltages, the resonance curve has a Lorentzian shape with a well-defined maximum [17]. The cantilever behaves like a linear oscillator with viscous damping, with the AFM tip stuck to the sample surface and following the surface motion. Above a critical surface shear ultrasonic vibration amplitude, typically 0.2 nm, the amplitude maximum of the resonance curves does not increase further, and the shape of the resonance curves change indicating the onset of sliding friction [17]. The information obtained from the analysis of the resonance curves in Fig. 12 supports the interpretation of torsional MD-UFFM curves discussed in section 3.1. In the MD-UFFM responses in Fig. 10, two different regimes are also distinguished. At low shear excitation voltages, the lateral mechanical diode effect leads to an increasing lateral ultrasonic force due to increasing shear vibration amplitude. Above a critical surface shear ultrasonic vibration amplitude, a maximum ultrasonic force is reached, and sliding begins.

In TR-AFM [66-71] torsional vibrations of the cantilever are excited via two piezoelectric elements mounted beneath the holder of the chip, which vibrate out-of-phase, in such a way that they generate a rotation at the long axis of the cantilever. The TR-mode can be implemented in contact, near-contact and non-contact modes, and provides information about surface shear elasticity, viscoelasticity and friction. When operating in contact, torsional cantilever resonance curves such as those in Fig. 12 have also been observed [68]. In the TR mode, the torsional resonance amplitude (or phase) can be used to control the feedback loop and maintain the tip/sample relative position through lateral interaction. Frequency modulation procedures have also been implemented for TR-AFM measurements [72].

### 3.4 Information from MD-UFFM data

*MD-UFFM curves:*

As in UFM, in MD-UFFM the data are typically collected in the form of *ultrasonic curves*, obtained by monitoring the mechanical-diode cantilever responses as a function of the shear ultrasonic excitation amplitude.

As discussed in section 3.1, the torsional MD-UFFM response stems from the lateral time-averaged force exerted upon a cantilever tip in contact with a sample surface, and scanning laterally over the surface at low typical AFM velocities when shear ultrasonic vibration of sufficiently high amplitude is excited at the tip-sample contact. Properties such as shear contact stiffness, shear strength and friction of surfaces at a nanometer scale are obtained in lateral force microscopy (LFM), also named Friction Force Microscopy (FFM) [62, 73]. In MD-UFFM, the excitation of shear ultrasonic vibration at the tip-sample contact leads to relative tip-surface velocities of mm s$^{-1}$ or larger, and the evaluation of these properties in these different experimental conditions may bring additional light to the understanding and control of nanoscale friction. Also, it is expected that MD-UFFM will provide subsurface information related to subsurface inhomogeities.

In the realm of continuum mechanics, for a sphere-plane geometry, the lateral stiffness of a contact is given by [74]:

$$K_{contact} = 8a_c G^* \qquad (eq.6)$$

where $a_c$ is the contact radius, and $G^*$ is the reduced shear modulus, defined as:

$$\frac{1}{G^*} = \frac{2-\nu_t}{G_t} + \frac{2-\nu_s}{G_s} \qquad (eq.7)$$

being $G_t$, $G_s$, $\nu_t$, $\nu_s$ the shear moduli and the Poisson´s ratios of the tip and the sample, respectively. This equation is valid for various continuum elasticity models and does not depend on the interaction forces. For small displacements it is reasonable to assume that there is no change in the contact area.

The elastic response of the tip-sample contact in shear can be described by a series of springs. A lateral displacement of the sample $\Delta z$ is distributed between three springs:

$$\Delta x = \Delta x_{contact} + \Delta x_{tip} + \Delta x_{cantilever} \qquad (eq.\,8)$$

And the lateral force $F_{lat}$ at the contact is given by

$$F_{lat} = k_{eff}\Delta x \qquad (eq.\,9)$$

being $K_{eff}$ an effective contact stiffness

$$\frac{1}{K_{eff}} = \frac{1}{K_{contact}} + \frac{1}{K_{tip}} + \frac{1}{c_L} \qquad (eq.\,10)$$

where $K_{contact}$ is the lateral contact stiffness, $K_{tip}$ is the lateral elastic stiffness of the tip, and $c_L$ is the lateral spring constant of the cantilever, considered as a point mass. For most commercial cantilevers, only the torsional spring constant is relevant for the estimation of $c_L$. In FFM experiments, the lateral stiffness of the tip is comparable or even smaller than the lateral stiffness of the cantilever [75].

For larger displacements at the contact, the threshold force to overcome the static friction is reached, and the tip starts to move. In FFM, $K_{eff}$ can be measured from the so-called friction force loops, lateral force vs. lateral position, in which a sticking part where the tip essentially stays at the same position and a sliding part can be easily distinguished. $K_{eff}$ is given by the slope of the sticking part.

The shear strength can be defined as:

$$F_f = \tau A = \tau \pi a_c^2 \qquad (eq.\,11)$$

where $F_f$ is the friction force, and $A$ is the contact area. From eq. 11, eq. 6 and eq. 7 we obtain an expresion for the shear strength, independent on the contact diameter $a_c$.

$$\tau = \frac{64 G^{*2} F_f}{\pi (K_{contact})^2} \qquad (eq.\,12)$$

It is well known from FFM studies that at typical low AFM scanning velocities, nanoscale friction proceeds by stick-slip. Once static friction at the tip-sample contact is overcome, the tip "slips" to a next static position and "sticks" there until the surface displacement is again large enough so that a threshold force needed for it to slip is reached again. Stick-slip also occurs at the micro and macro scales and can be observed whatever the chemical nature of the solids in contacts, and the state of their surfaces provided that the loading system is soft enough. Stick-slip friction with atomic periodicity has been demonstrated in numerous LFM experiments with atomic resolution, in which the lateral force exhibits a periodic, sawtoothlike behaviour [62]. According to the Tomlinson model, the tip is considered to move in the periodic potential field formed by the subtrate lattice while being dragged along the surface by means of spring-type interactions. Atomic-scale stick-slip is usually limited to low load regime, and sharp tips, although, atomic-scale stick slip at high loads have also been observed. The latter may be restricted to layered materials or to the presence of some lubricating contamination films. In the Prandtl-Tomlinson model, the total potential experienced by the tip is given by:

$$V_{tot}(x,t) = -\frac{E_o}{2} \cos\frac{2\pi x}{a} + \frac{1}{2} K_{eff}(x - vt)^2 \qquad (eq.\,13)$$

where $E_o$ is the peak-to-eak amplitude of the surface potential, $a$ is the lattice constant of the surface, K*eff* is the effective lateral spring constant and *v* is the velocity of the sample.

The model for MD-UFFM described in section 3.2 is based on the Tomlinson model This accounts qualitatively quite well for the experimental results (see Fig. 11 and related text). In principle, the application of this model allows us to obtain $K_{eff}$, $F_f$ and $\tau$ from MD-UFFM data, and also learn about the relationship of these magnitudes with the surface lateral potential, its amplitude $E_o$ and periodicity $a$, and the mechanisms of friction in the presence of shear ultrasonic vibration at the tip-sample contact. In FFM atomic-scale stick-slip friction experiments performed at low loads, the values obtained for $K_{eff}$ suggest that the area of contact consists of just a few atoms, precluding the application of continuum mechanical models in those cases.

Fig. 13 shows MD-UFFM responses on Si(111) recorded at different normal set-point forces, including the torsion curves recorded in both forward and backward scans. For higher normal loads, the magnitude of torsional MD signal increases, and a higher critical shear ultrasonic amplitude is required to reach the flat torsion regime attributed to sliding. These results are also in agreement with the model sketched in Fig. 11. For higher loads, the magnitude of the surface interatomic potential is expected to be larger [76].

In Fig. 13, the distance between the torsion curves recorded in forward and backward scans is proportional to the magnitude of the friction force. The results indicate that friction reduces as a result of the excitation of shear ultrasonic vibration at the tip-sample contact, and that in this case friction vanishes in the flat MD torsional response regime. Physically, the onset of a lateral ultrasonic force is necessarily related to a reduction of friction (see Fig. 11). The effect might be related to the observations in ref. [77]. There it was concluded that a cantilever may exhibit apparent stick-slip motion, and hence reveal a nonzero mean friction force, even when the tip-surface contact is completely thermally lubricated by fast activated jumps of the tip apex, back and forth between the surface potential wells. Even though, as mentioned before, in MD-UFFM, the excitation of shear ultrasonic vibration at the tip-sample contact leads there to relative tip-surface velocities of the order of mm·s$^{-1}$ or larger within the contact, it is still the displacement of the position of the cantilever center of mass relative to the surface which determines the contact velocity.

The lift-off (deflection) signals that accompany the MD torsional response in Fig. 13 has been attributed to the presence of an ultrathin viscous liquid layer at the tip-sample contact which develops hydrodynamic pressure when sheared at ultrasonic velocities [15]. The shape of those lift-off curves is essentially different from the typical UFM MD deflection response that results from the excitation of normal ultrasonic vibration [38]. In the MD-UFFM case, the cantilever deflection increases linearly as the shear ultrasonic vibration amplitude is increased, and no apparent jump-off is noticeable. Slight deviations of the linear shape of the deflection curve when the maximum deviation of the initial cantilever torsion is reached may be related to a coupling of the cantilever lateral and vertical motions at the onset of the sliding regime. The presence of a squeezed liquid layer at the Si surface - Si tip contact has been previously considered to explain a reduction of friction in ambient conditions as a result of the excitation of normal ultrasonic vibration at amplitudes not sufficiently large to break the tip-sample contact during the ultrasonic period [47]. However, such a lift-off has not been observed when performing MD-UFFM experiments on Si in liquid environment [37]. Fig. 14 (a) (b) shows lateral mechanical diode responses – MD-UFFM signals – measured on silicon, in milliQ water. The torsion MD-UFFM curves in liquid are similar as in air, although in liquid environment they appear considerably noisier [37]. Any lift-off MD-UFFM deflexion signal has been observed in MD-UFFM experiments performed on highly oriented pyrolitic graphite (HOPG) in air, either [78].

*MD-UFFM images*

As demonstrated in Fig. 15, MD-UFFM can also be implemented in an imaging mode, using a lock-in amplifier to monitor the signal at the amplitude modulation frequency. Fig. 15 shows FFM images in forward (a), and backward (b) scans, an ultrasonic MD-UFFM torsion image (c) and ultrasonic MD-UFFM curves recorded at different points on the same surface region. As in UFFM [15], MD-UFFM images are independent of the scanning direction, i.e. not influenced by topography-induced lateral forces. Whereas Fig. 15 evidences the possibility to map surface properties in MD-UFFM, a precise interpretation of the MD-UFFM contrast in Fig. 15 is nevertheless, not straightforward, and deserves further investigations.

Summarizing, MD-UFFM appears as an interesting new technique, based in the study of the lateral mechanical diode cantilever response in the presence of shear surface ultrasonic vibration. Although in a very incipient state of development, the technique show promise of being useful for the measurement of shear elasticity, shear strength and friction at the nanometer scale, to probe the surface interatomic potential, for investigation of the atomistic mechanisms involved in nanoscale tribology, the study of elastohydrodynamic lubrication effects in confined layers at nanogaps, for the characterization of boundary lubricants, etc.

## 4. Heterodyne Force Microscopy (HFM): Beats at Nanocontacts.

### 4.1 Beats at nanocontacts.

If at a nanocontact, we excite vibration of frequency $\omega_1$ at one end, and vibration of frequency $\omega_2$ at the other end, being the excitation frequencies different but close to each other ($\omega_1 \neq \omega_2$; $\omega_1 \approx \omega_2$), the separation between both ends $d$ will vary periodically with time, one cycle of this variation including many cycles of the basic vibrations at both ends, and with a frequency equal to the average of the two combining frequencies. The phenomena is actually the description of a beating effect [71] applied to the nanocontact. If $y_1$ and $y_2$ are the positions of each nanocontact end,

$$y_1(t) = A\sin\omega_1 t \qquad (eq.\,14)$$

$$y_2(t) = d_o + A\sin\omega_2 t \qquad (eq.\,15)$$

$d_o$ its separation in the absence of vibration, and $A$ the vibration amplitude of each end, then, the separation of both ends will vary with time according to:

$$d(t) = y_2(t) - y_1(t) = d_o + 2A\sin\left(\frac{\omega_2 - \omega_1}{2}t\right)\cos\left(\frac{\omega_2 + \omega_1}{2}t\right) \quad (eq.\,16)$$

Eq. 16 holds in fact for any values of $\omega_1$ and $\omega_2$, but the description of the beat phenomenon is physically meaningful only if $|\omega_2 - \omega_1| \ll \omega_2 + \omega_1$. Then, over a substantial number of cycles, the vibration approximates to sinusoidal vibration with constant amplitude and with frequency $(\omega_2+\omega_1)/2$.

The term $\cos\left(\frac{\omega_2+\omega_1}{2}t\right)$ describes the rapidly oscillating factor in eq. 16, and will always lie between the limits ±1. The distance between the two ends in the nanocontact will vary between minimum $d_o-2A$ and maximum $d_o+2A$ values at a frequency given by $|\omega_2 - \omega_1|$, i.e. at the beat frequency.

If we consider now that the nanocontact is this formed by the tip of an AFM cantilever and a sample surface (see Fig. 16), the beat effect implies that the tip-sample distance varies between a minimum value and a maximum value at the beat frequency in the case we simultaneously excite ultrasonic vibration at the tip and the sample surface at slightly different frequencies. Notice that the beat frequency is in fact much smaller than the actual tip and sample vibration frequencies. Hence, if the tip-sample distance variation in the beats is such that the tip-sample force remains in the linear regime, if we try to detect the force that acts upon the cantilever at the beat frequency, we will find that the tip-sample distance, and hence the force upon the cantilever, is varying from the minimum value to the maximum value many times in the time scale that we will use to track the beat frequency, and we will only be able to detect the averaged value of this force, which will be null in the linear case. However, if the variation of tip-sample distance during a beat cycle is such that sweeps the nonlinear tip-sample force regime, when trying to measure the force acting upon the cantilever at the beat frequency, we will detect the averaged force, which will change following the periodicity of the beats.

In Heterodyne Force Microscopy (HFM) [27], ultrasound is excited both at the tip (from a transducer at the cantilever base) and at the sample surface (from a transducer at the back of the sample) at adjacent frequencies, and mixed at the tip-sample gap. If $d_o$ is the initial tip-sample indentation, and the vibration amplitude A of the tip and the surface is the same, the tip-sample force will vary according to eq. 16, assuming that for instance $\omega_1$ corresponds to the frequency vibration of the sample, and $\omega_2$ to the frequency vibration of the tip. In HFM, the modulation frequency is usually chosen much lower than the first cantilever resonance frequency. The cantilever will not be able to follow the force exerted at the frequency $(\omega_1+\omega_2)/2$ due to its inertia. However, provided that the low-frequency varying tip-sample separation is large enough to cover the nonlinear range of the tip-sample interaction force, an ultrasonic force (stronger for larger amplitudes) will act upon the cantilever and displace it from its initial position. Owing to the varying ultrasonic force, the cantilever vibrates at the difference mixed frequency. In principle, even if the modulation frequency is chosen higher than the first cantilever resonance [30] or coincident with a cantilever contact resonance [31] the beat effect should also lead to the activation of an ultrasonic force at the beat frequency, provided that the tip-sample distance is varied over the nonlinear tip-sample force regime as a result of the tip and sample high frequency vibration. Also, the effect should similarly work if the cantilever is operated in a dynamic AFM mode.

An important feature of the beat effect is that it facilitates the monitoring of phase shifts between tip and sample ultrasonic vibrations with an extremely high temporal sensitivity. In HFM, it has been demonstrated that small differences in the sample dynamic viscoelastic and/or adhesion response to the tip interaction result in a shift in phase of the beat signal that is easily monitored. In this way, HFM makes possible to study dynamic relaxation processes in nanometre volumes with a time-sensitivity of nanoseconds or even better.

## 4.2 Experimental implementation of HFM

The experimental set-up for HFM is shown in Fig. 17. The technique can be implemented by appropriately modifying a commercial AFM equipment [27]. For HFM, PZT ceramic piezos are attached to the sample and the tip holder. Both the samlple and the cantilever are bonded to the corresponding piezos using a thin layer of crystalline salol (phenyl salicilate). Two function generators are needed to simultaneously excite sinusoidal vibration of the sample surface and the cantilever tip at two adjacent ultrasonic frequencies. In [27], sample and tip vibrations were excited at frequencies in the MHz range, differing in some KHz. The synchronous signals from both generators at the high-frequency excitation can be electronically mixed using a simple electronic mixer, which provide as an output a reference signal at the difference frequency. By means of the lock-in amplifier, the vibration of the cantilever at the beat frequency, i.e. the HFM signal in this case, can be easily monitored in amplitude and phase.

The recently proposed technique of Scanning Near-Field Ultrasound Holography (SNFUH) [30] is implemented in a similar way as HFM, choosing a difference frequency (beat frequency) in the range of hundreds of KHz, above the first cantilever resonance frequency. In Resonant Difference Frequency Atomic Force Ultrasonic Microscopy (RDF-AFUM) [31], the difference frequency (beat frequency) is chosen to be coincident with a high-order cantilever contact resonance.

## 4.3 Comparison of HFM with UFM

If in UFM the surface ultrasonic vibration excited from a piezo located at the back of the sample is modulated in amplitude using a sinusoidal shape instead of the customary triangular or trapezoidal modulation shape, the tip-sample distance will vary similarly is it does in the case of HFM (see Fig. 16). Actually, in UFM we could also collect an Amplitude-UFM and a Phase-UFM signals using the lock-in amplifier, although up-to-now usually only the Amplitude-UFM response has been considered. The main important difference between UFM and HFM lies in the fact that in UFM the ultrasonic vibration is input into the system only from one end of the tip-sample nanocontact, while in HFM, both nanocontact ends are independently excited.

So far the excitation from one end of the nanocontact will be transmitted through the contact to the other end, i.e. ultrasonic vibration from the sample surface will propagate to the AFM cantilever tip, Amplitude-HFM and UFM signals are expected to be quite similar. In fact, as we mentioned in section 2.2, UFM can also be implemented in the so-called Waveguide UFM mode, in which the ultrasonic vibration at the tip-sample contact is excited from a piezo located the cantilever base, and no significant qualitative differences in the UFM response have been encountered when comparing ultrasonic curves received in either case [36]. The comparison of UFM and waveguide UFM studies on a same sample is interesting in order to differentiate surface from subsurface effects. In HFM, this same kind of information may be available by appropriate modification of the sample or tip ultrasonic vibration amplitudes. In any case, for some studies, the use of a triangular or trapezoidal shape for ultrasonic amplitude modulation may be preferred, and UFM may still be the technique of choice.

The great strength of HFM versus UFM relies on the phase measurements. By monitoring the phase of the cantilever vibration at the beat frequency, HFM allow us to detect slight changes in phase of the sample vibration with time resolution of fractions of the sample and cantilever ultrasonic periods. If the excitation frequencies are in the MHz regime, and the difference frequency is of some KHz, phase delays between tip and sample vibrations of the order of nanoseconds are easily detectable [27]. Notice that even though it is possible to perform phase-UFM by monitoring the phase of the cantilever vibration at the ultrasonic modulation frequency because of the mechanical diode response, in the absence of forced ultrasonic excitation of the tip, the phase differences between sample and tip ultrasonic vibrations cannot be straightforwardly measured, and the time-sensitivity to phase-delay-related processes is in the best of cases limited to the ultrasonic period, at least easily three orders of magnitude smaller than in the HFM case.

**4.4 Information from HFM: time resolution**

As discussed in section 4.3, the big potential of HFM is based on its capability to perform phase-delay measurements with an extremely high sensitivity. Phase delays may originate from different elastic or viscoelastic properties, from different in-depth locations of a same-type of elastic inhomogeneity, and in general from any local dissipative process activated by mechanical vibration. So far ph-HFM provides a means to probe a local response in an extremely short time, the technique may reveal dissipation due to extremely quick transitions, otherwise unresolved from other dissipative effects occurring at larger time scales. Phase-HFN has been applied to PMMA/rubber nanocomposites that consist in an acrylic matrix, a copolymer based upon PMMA, and toughening particles, composed of a core of acrylic enclosed with rubber with a bonded acrylic outer shell to ensure good bonding to the matrix [27] (see Fig. 18). Using Phase-HFM, it has been possible to distinguish differences in contrast at identical thin polymer layers with different boundary constraints on the nanometer scale. In the Ph-HFM images a different viscoelastic and/or adhesion hysteresis response time of the PMMA on top of the rubber that is not linked to the PMMA rubber matrix is clearly distinguish. Such different PMMA responses cannot however be appreciated from the Amplitude-HFM images.

Using the recently proposed SNFUH mode [30], perform similarly as phase-HFM, elastic information of buried features have been obtained from phase measurements with great sensitivity. In the RDF-AFUM procedure, subsurface nanoscale elastic variation have also been observed [31]. In RDF-AFUM the beat effect is used as in HFM, but the beat frequency is chosen to be coincident with a cantilever contact resonance. In ref. [31], an analytical model is proposed to account for the RDF-AFUM response, considering the interaction of the ultrasonic wave generated at the bottom of the sample with nano-/microstructural features within the sample bulk material, and the nonlinear cantilever tip-sample surface interactions.

Nevertheless, up-to-date, the data reported with beat-effect related AFM techniques is still very limited. The beat effect may facilitate opportunities ranging from the precise evaluation of elastic or viscoelastic response of nanostructures, the analysis of snap shots or transient states in the mechanical response of nanoobjects, the implementation of nanoscale time-of-flight experiments with high temporal resolution, or the quick transmission of information through nanocontacts by mechanical means. The use of higher beat frequencies opens up the possibility to scan at higher lateral scanning speeds while recording material information. Phase-HFM facilitates straightforward measurements of phase-delays between tip and sample vibrations, with extremely high sensitivity. The opportunities bring about by this technique are still to explore.

**Acknowledgments**

M. T. C. thanks S. K. Biswas and W. Arnold for carefully reviewing this chapter. Financial support from the Junta de Comunidades de Castilla-La Mancha (JCCM) under project PBI-08-092 is gratefully acknowledged.

33[28] *Detection of nanomechanical vibrations by dynamic force microscopy in higher cantilever eigenmodes*, A. San Paulo, J. P. Black, R. M. White, and J. Bokor, Appl. Phys. Lett. 91 (2007) 053116.

[29] *Ultrasonic Machining at the Nanometer Scale*, M. T. Cuberes, J. of Phys: Conf. Ser. 61 (2007) 219.

[30] *Nanoscale imaging of buried structures via Scanning Near-Field Ultrasound Holography*, G. S. Shekhawat and V. P. Dravid, Science 310 (2005) 89.

[31] *Nanoscale subsurface imaging via resonant difference-frequency atomic force ultrasonic microscopy*, S. A. Cantrell, J. H. Cantrell, and P. T. Lillehei, J. Appl. Phys. (2007) 114324m.

[32] *Nanoscale determination of phase velocity by scanning acoustic force microscopy*, E. Chilla, T. Hesjedal, and H. –J. Fröhlich, Phys. Rev. B 55 (1997) 15852.

[33] *Imaging of acoustic fields in bulk acoustic-wave thin-film resonators*, H. Safar, R. N. Keliman, B. P. Barber, P. L. Gammel, J. Pastalan, H. Huggins, L. Fetter, and R. Miller, Appl. Phys. Lett. 77 (2000) 136.

[34] *Mapping surface elastic properties of stiff and compliant materials on the nanoscale using ultrasonic force microscopy*, F. Dinelli, M. R. Castell, D. A. Ritchie, N. J. Mason, G. A. D. Briggs, and O. V. Kolosov, Philosophical Mag. A 80 (2000) 2299.

[35] *Waveguide ultrasonic force microscopy at 60 MHz*, K. Inagaki, O. Kolosov, A. Briggs, and O. Wright, Appl. Phys. Lett. 76 (2000) 1836.

[36] *Nonlinear detection of ultrasonic vibration of AFM cantilevers in and out of contact with the sample*, M. T. Cuberes, G. A. D. Briggs, and O. Kolosov, Nanotechnology 12 (2001) 53.

[37] *Nanoscale ultrasonics in liquid environments,* M. T. Cuberes, J. of Physics: Conf. Ser. (in press).

[38] *Measurements of stiff-material compliance on the nanoscale using ultrasonic force microscopy*, F. Dinelli, S. K. Biswas, G. A. D. Briggs, and O. V. Kolosov, Phys. Rev. B 61 (2000) 13995.

[39] *AFM and Acoustics: Fast, Quantitative Nanomechanical Mapping*, B. D. Huey, Annu. Rev. Mater. Res. 37 (2007) 351.

[40] *Hysteresis of the cantilever shift in ultrasonic force microscopy*, K. Inagaki, O. Matsuda, and O. B. Wright, Appl. Phys. Lett. 80 (2002) 2386.

**FIGURE CAPTIONS**

Fig. 1   SAM operating in reflexion mode (from ref. [7])

Fig. 2   Set-up for *(a)* Scanning Transmision Acoustic Microscopy and *(b)* Scanning Acoustic Holography (from ref. [9])

Fig. 3   Detection of surface out-of-plane ultrasonic vibration with the tip of an AFM cantilever via the mechanical-diode effect *(a)* When the surface vibration amplitude is sufficiently high, tip experiences an *ultrasonic force $F_{us}$*. *(b)* Tip-sample force $F_{t-s}$ versus tip-sample distance *d* curve.

Fig. 4   Set-up for UFM measurements

Fig. 5   *(a)* Schematic plot of a force-indentation curve *(b)* Schematic ultrasonic cantilever deflection (mechanical-diode signal) induced by out-of-plane sample vibration of increasing amplitude (from ref. [38])

Fig. 6   Experimental UFM curves. Cantilever deflexion $z_c$ recorded for different tip-sample forces $F_n$ on highly oriented pyrolitic graphite (HOPG).

Fig. 7   Modified force-indentation curves (retraction branches) for a JKR solid-solid (tip-sample) contact in the presence of normal ultrasonic vibration of different amplitudes $a_i$. The black line is the force exerted by the cantilever considered as a point mass (from ref. [38])

Fig. 8   Machining of nanotrenches and holes on silicon using a UFM (from [59])

Fig. 9   (*a*) Topography on the HOPG surface (700 × 700) nm *(b) (c)* Ultrasonic-AFM images recorded in sequence over nearly the same surface region as in (a). A subsurface dislocation not noticeable in the topographic image is enclosed by the ellipse in (b) and (c). In (c) the dislocation is laterally displaced. (From ref. [49])

Fig. 10  Experimental evidence of the lateral MD effect (see text) (from ref. [26])

Fig. 11  Physical model for MD-UFFM The surface atoms are laterally displaced due to the shear surface vibration, but due to its inertia, the cantilever cannot follow the surface lateral displacements at ultrasonic frequencies not coincident with a torsional cantilever resonance (from ref. [27])

Fig. 12  Torsional vibration amplitude of the cantilever as a function of the excitation frequency. Measurements on bare silicon. The different curves correspond to increasing excitation voltages applied to the shear-wave piezotransducer (from ref. [17])

Fig. 13  MD-UFFM reponses on Si(111) for different normal loads (from ref. [26])

Fig. 14  MD-UFFM on Si, in milliQ water (from ref. [37])

Fig. 15  Octadecylamine on mica. *(a) (b)* FFM images in the absence of ultrasound. *(c)* MD-UFFM image on the same surface region. *(d)* Torsional MD-UFFM curves on different surface points measured while recording (c).

Fig. 16  Beats at the tip-sample contact

Fig. 17  Set-up for HFM (from ref. [27])

Fig. 18  HFM on PMMA/rubber nanocomposites (from ref. [27])

Fig. 1

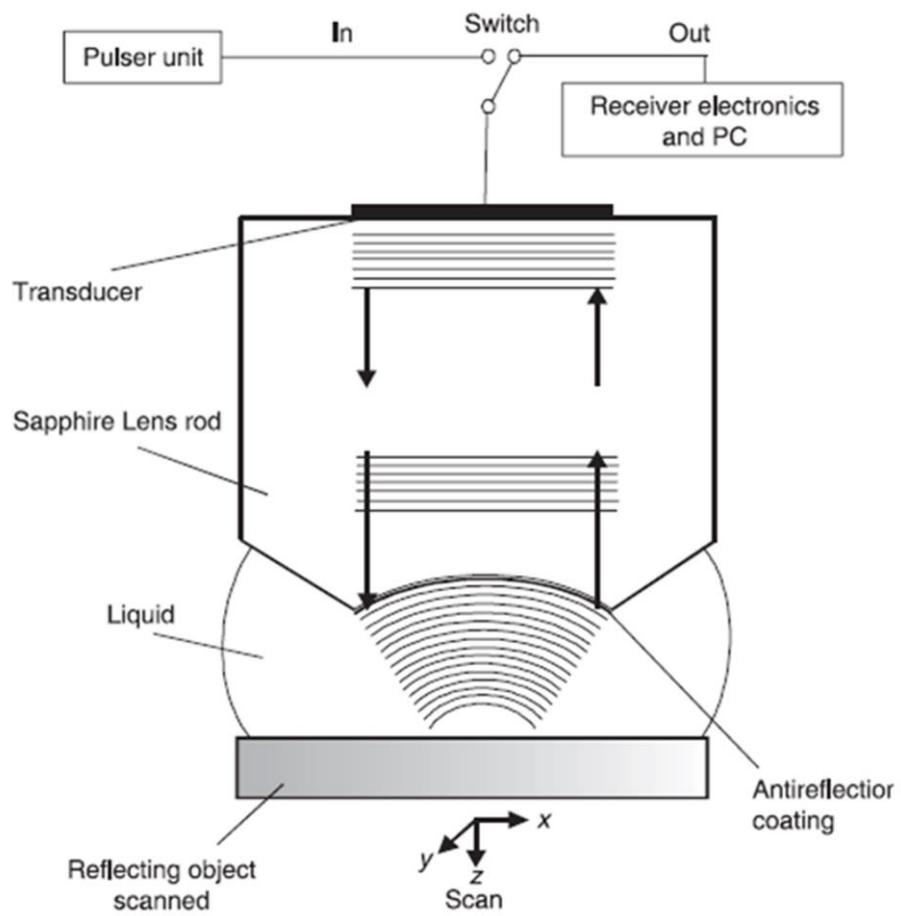

Fig. 2

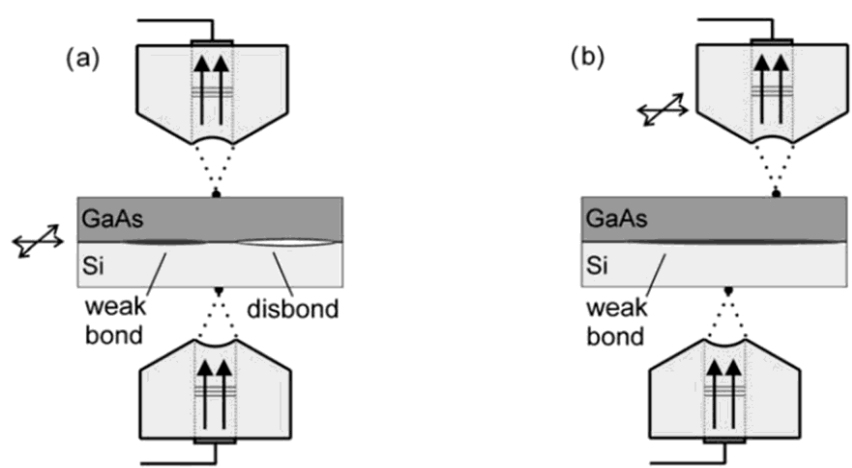

Fig. 3

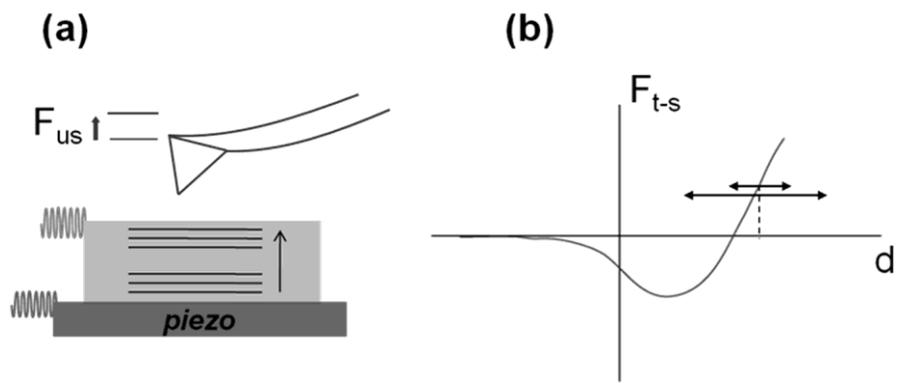

Fig. 4

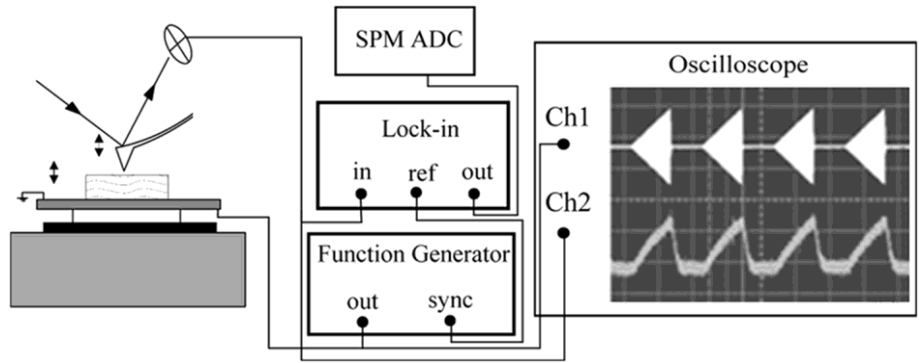

Fig. 5

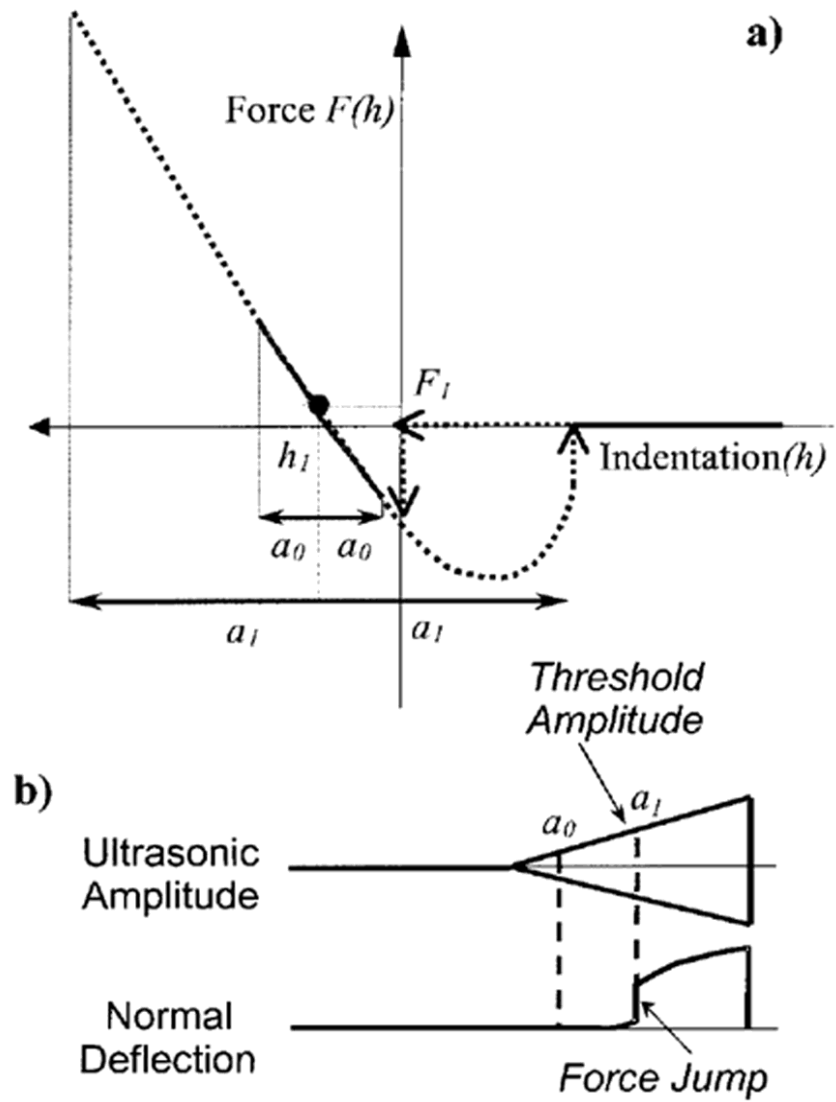

Fig. 6

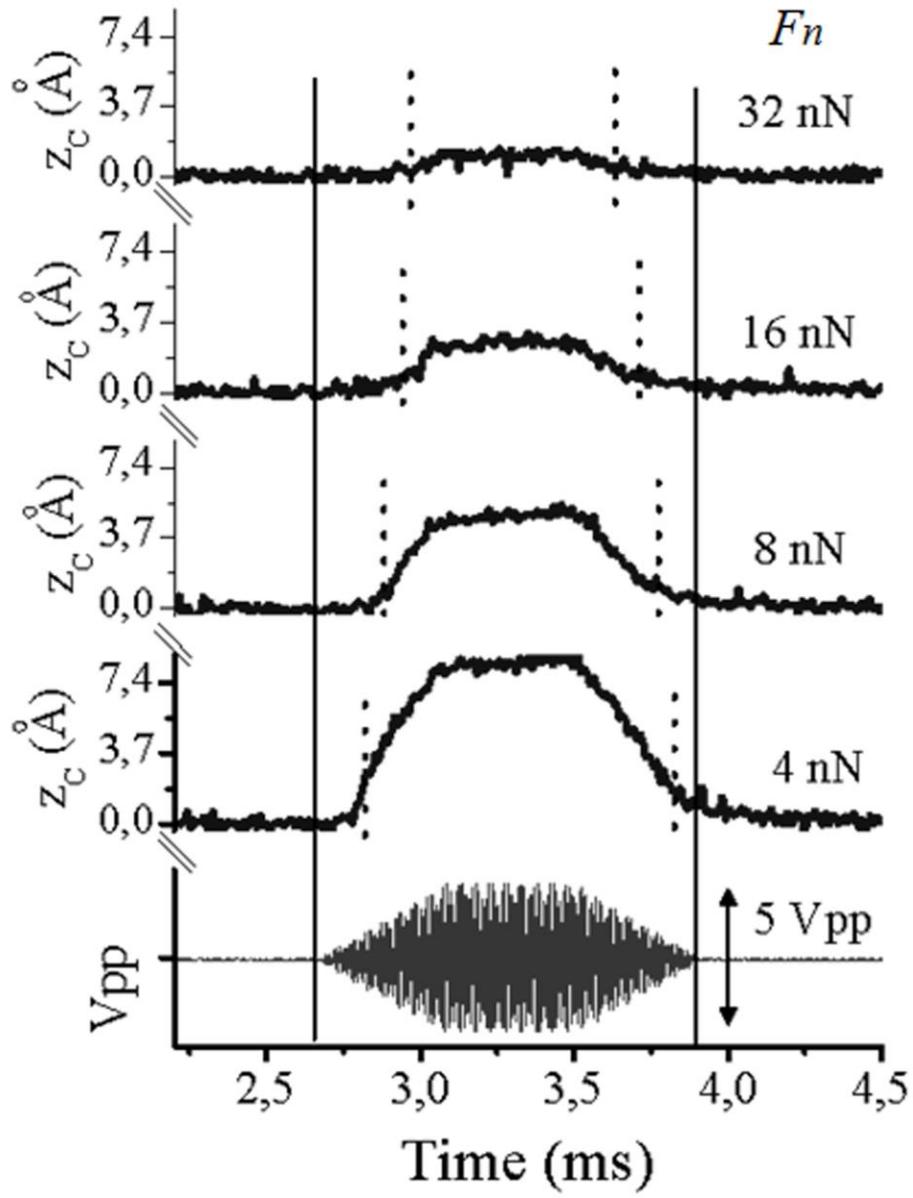

Fig. 7

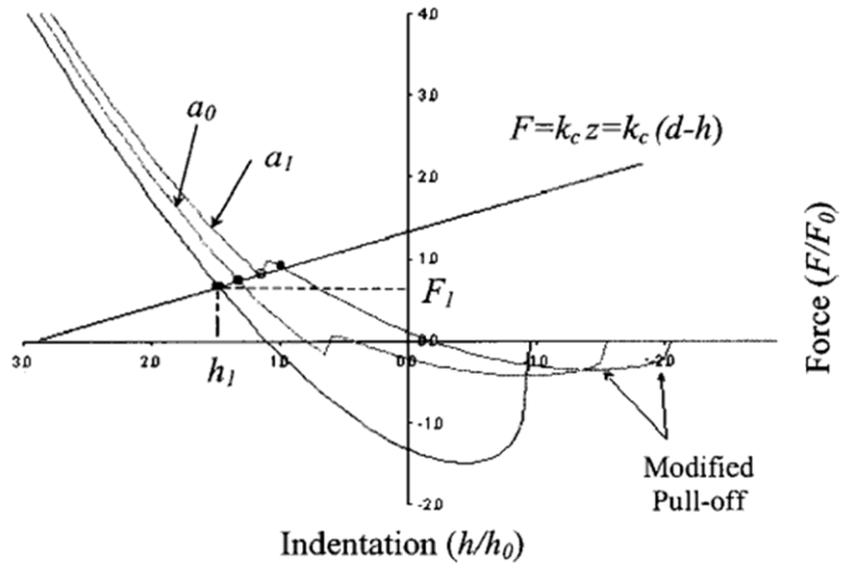

Fig. 8

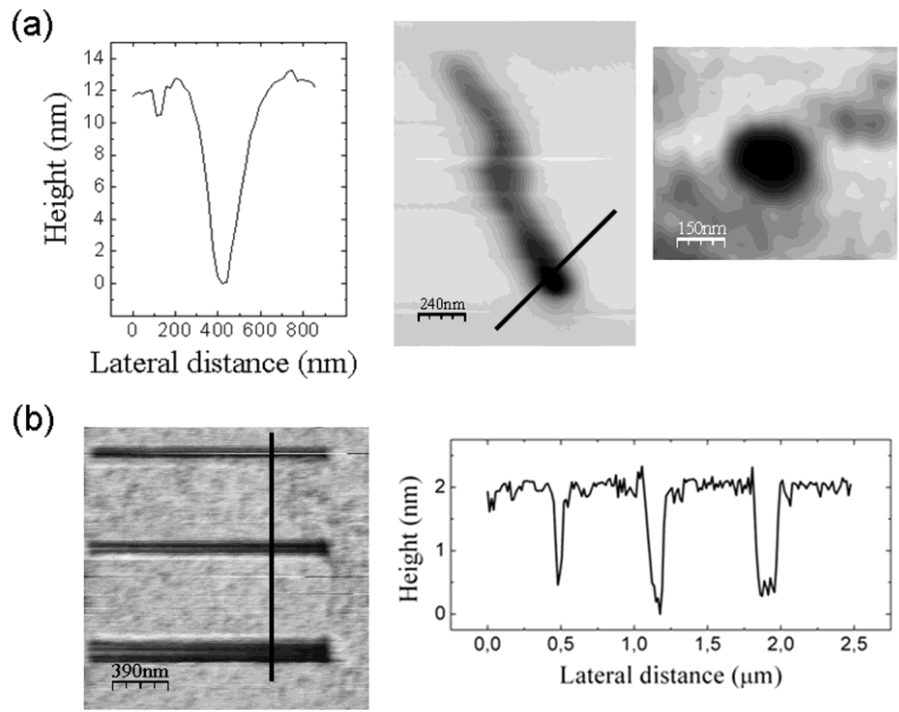

Fig. 9

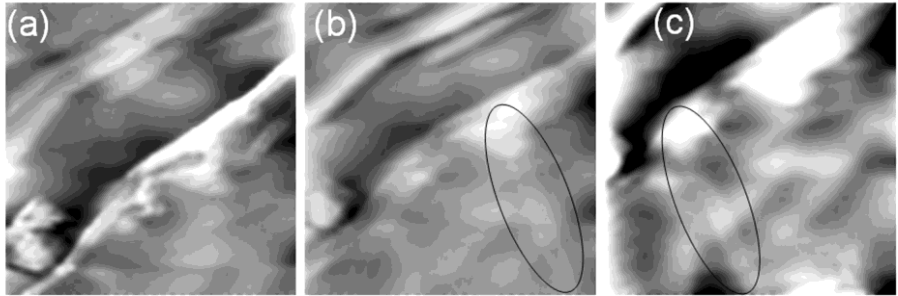

Fig. 10

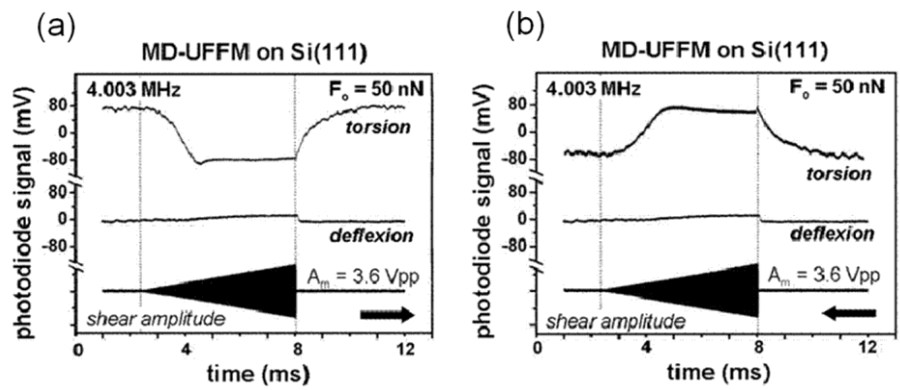

Fig. 11

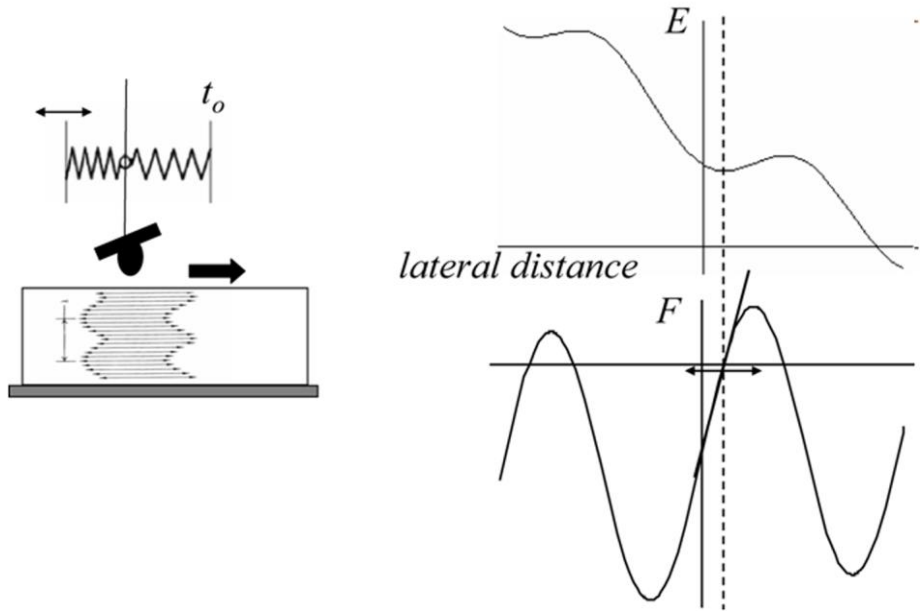

Fig. 12

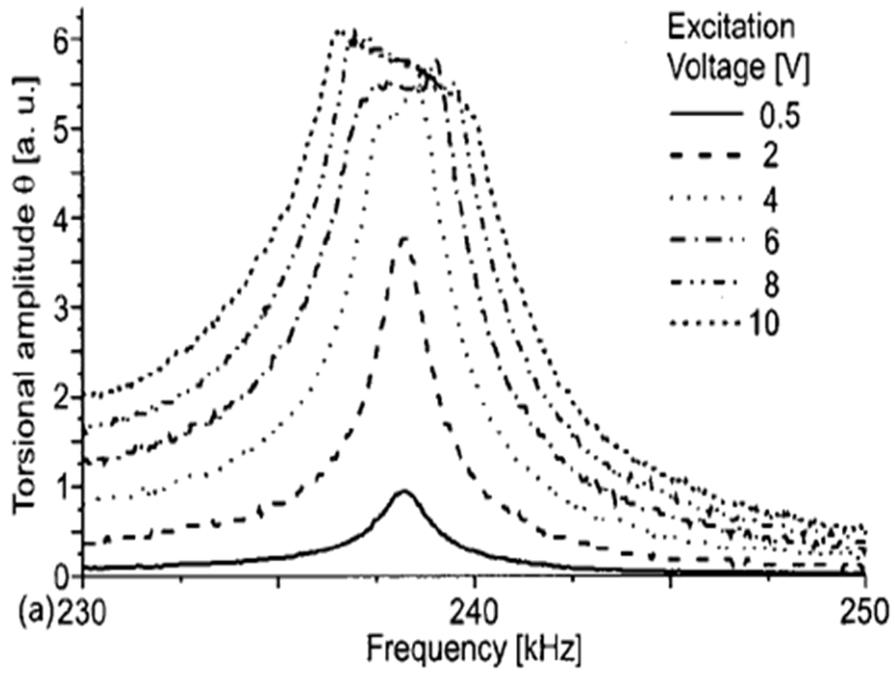

Fig. 13

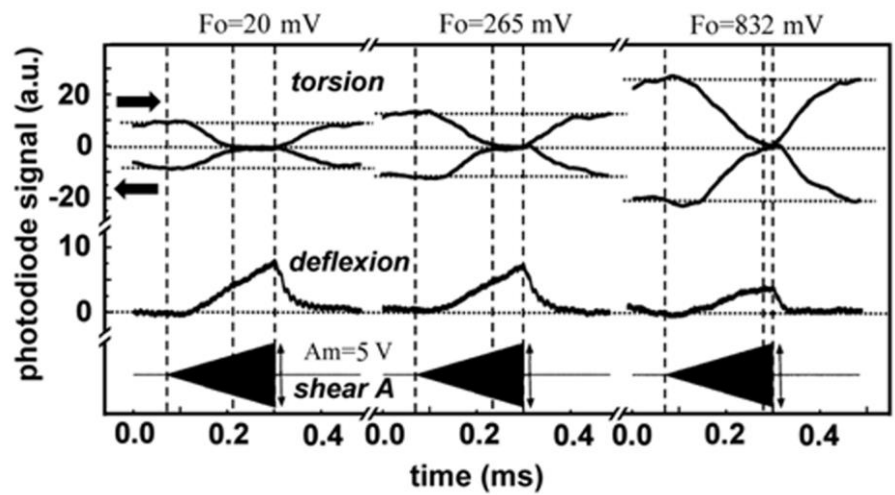

Fig. 14

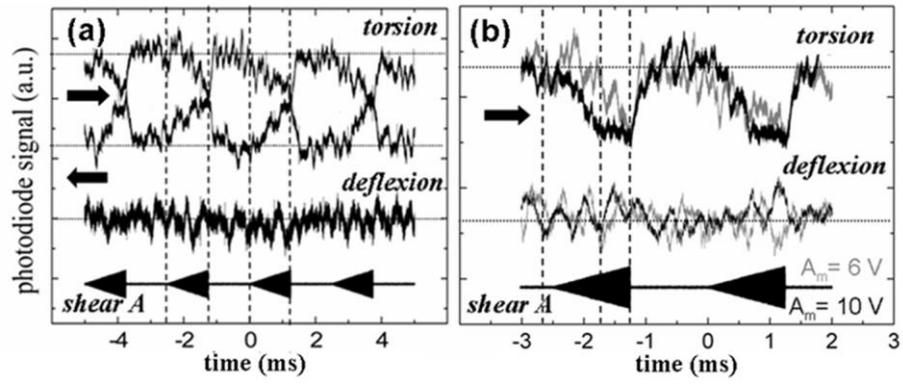

Fig. 15

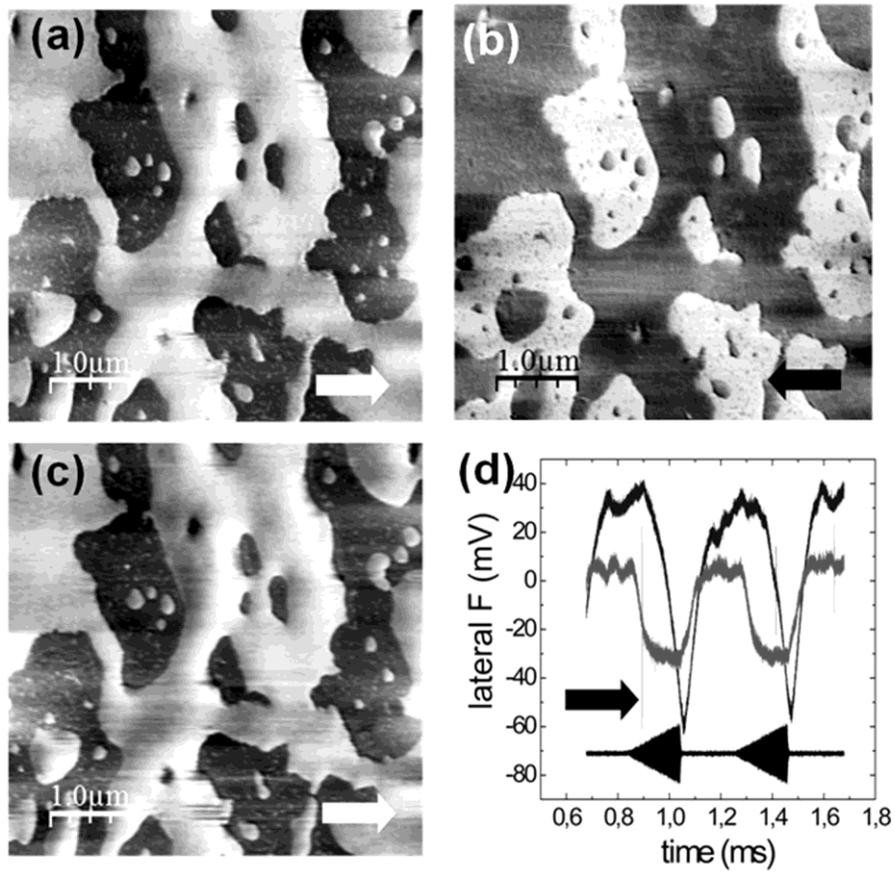

Fig. 16

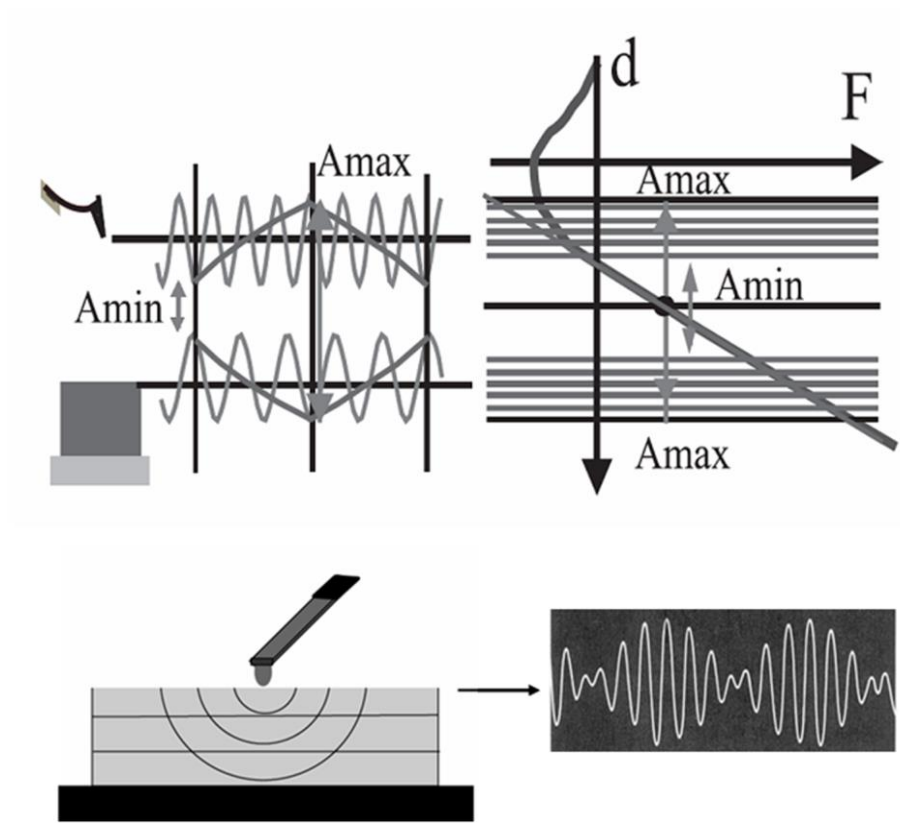

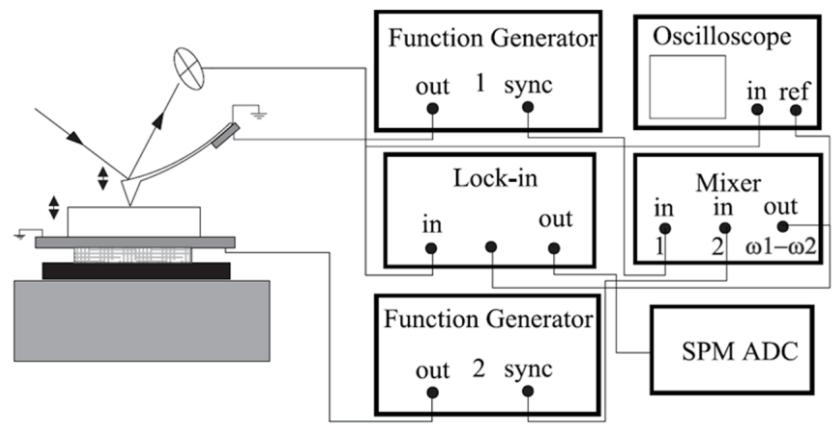

Fig. 17

Fig. 18

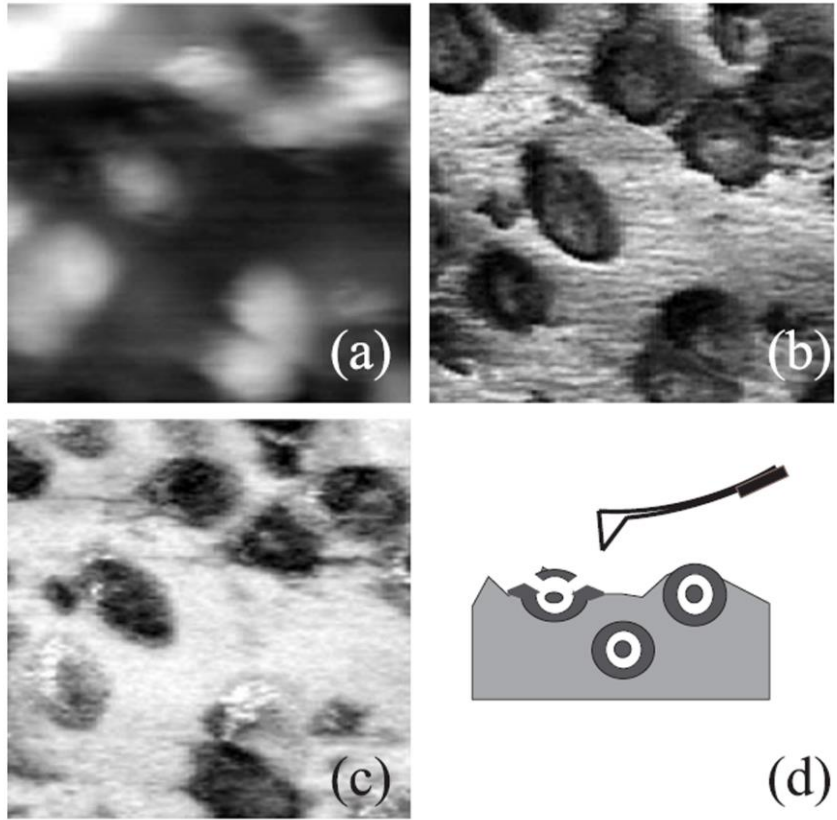